\documentclass[apj]{emulateapj}
\usepackage{apjfonts}
\RequirePackage{natbib}
\usepackage{lscape}
\def\teff{T$_{\rm eff}$}
\def\logg{log $g$} 

\begin{document}

\title{A Unique Star in the Outer Halo of the Milky Way \footnote{
The data presented herein were obtained at the W.M. Keck Observatory,
which is operated as a scientific partnership among the California
Institute of Technology, the University of California and the
National Aeronautics and Space Administration. The Observatory was
made possible by the generous financial support of the W.M. Keck
Foundation.}}

\author{David K. Lai{\altaffilmark{1,2}}, Constance M. Rockosi{\altaffilmark{1}}, Michael Bolte{\altaffilmark{1}}, Jennifer
  A. Johnson{\altaffilmark{3}},  Timothy
  C. Beers{\altaffilmark{4,5}}, Young Sun Lee{\altaffilmark{4,5}},
  Carlos Allende Prieto {\altaffilmark{6,7}}, Brian Yanny{\altaffilmark{8}}}
\altaffiltext{1}{UCO/Lick Observatory, Department of Astronomy and Astrophysics, University
  of California, Santa Cruz, CA 95064; david@ucolick.org, crockosi@ucolick.org
  bolte@ucolick.org.}
\altaffiltext{2}{NSF Astronomy and Astrophysics Postdoctoral Fellow}
\altaffiltext{3}{Department of Astronomy, Ohio State University, 140
  W. 18th Avenue, Columbus, OH 43210; jaj@astronomy.ohio-state.edu.}
\altaffiltext{4}{Department of Physics $\&$ Astronomy, CSCE (Center
  for the Study of Cosmic Evolution), Michigan State University, East
  Lansing, MI 48824, USA; lee@pa.msu.edu, beers@pa.msu.eduand}
\altaffiltext{5}{JINA (Joint Institute for Nuclear Astrophysics),
  Michigan State University, East Lansing, MI 48824, USA }
\altaffiltext{6}{
McDonald Observatory and Department of Astronomy, The University of
Texas, 1 University Station, C1400, Austin, TX 78712-0259, USA; callende@astro.as.utexas.edu}
\altaffiltext{7}{
Mullard Space Science Laboratory, University College London, Holmbury St.
Mary, Surrey RH5 6NT, United Kingdom
}
\altaffiltext{8}{
Fermi National Accelerator Laboratory, P.O. Box 500, Batavia, IL
60510; yanny@fnal.gov}

\begin{abstract}

As part of a program to measure abundance ratios in stars beyond 15 kpc
from the Galactic center, we have discovered a metal-poor star in the outer 
halo with a unique chemical signature. We originally identified it in
the Sloan Extension for Galactic Understanding and Exploration (SEGUE) 
survey as a distant metal-poor star. We obtained a follow-up spectrum using the 
Echellette Spectrograph and Imager (ESI) at the Keck 2 telescope, and measure 
[Fe/H] $=-3.17$, [Mg/Fe] =$-0.10$ and [Ca/Fe] $=+1.11$. This is one of the 
largest over-abundances of Ca measured in any star to date; the extremely
low value of [Mg/Ca]=$-1.21$ is entirely unique. To have found such an
unusual star in our small sample of 27 targets suggests that there may be previously
unobserved classes of stars yet to be found in situ in the Galactic halo. 

\end{abstract}


\section{Introduction}

The uniformity of certain abundance ratios in metal-poor stars has been
proven to be striking (e.g., \citealt{cayrel04,cohen04,arnone05,heres2,lai08}). This
uniformity is particularly evident in the $\alpha$-element abundance
ratios, indicating some combination of a well-mixed interstellar
medium (ISM) and common
star-formation history for nearby halo metal-poor stars. 
Exceptions exist (e.g., \citealt{ivans03,aoki07b,cohen07}), 
but these have been rarely seen in surveys to date.

The study of metal-poor stars, however, has been largely confined to
the solar neighborhood, and in effect the inner-halo region. Recent
work provides conclusive evidence that there is a population of stars at
Galactocentric distances greater than 15 kpc with kinematics,
stellar density profile and metallicity distribution distinct from
that of the inner halo \citep{carollo07}. The outer halo exhibits
a net retrograde motion and is on average composed of more metal-poor
stars than the inner halo. \citet{bell08} has found evidence for
substructure in this outer-halo region as compared to smooth stellar
distribution models. Analysis of nearby stars determined to be members
of the outer halo through kinematics (e.g., \citealt{roederer09}) have
shown an increased scatter in multiple abundance ratios relative to
stars that belong to the inner halo.

The formation of galaxies by hierarchical merging may provide a ready
explanation for these observations. \citet{bell08} showed that their
findings of substructure in the halo compared favorably with the
simulations of \citet{bullock05}, which assumed that merging and
redistribution of the member stars of satellite galaxies formed the
halo. Simulations by \citet{delucia08} found a more metal-poor outer
halo caused by dynamical friction sending the more massive (and
therefore more metal-rich) satellites into the inner halo, leaving the
stars of less massive satellites preferentially in the outer
halo. Depending on when these later satellites accreted to form the
outer halo, they will leave a signature in both the bulk metallicity
and $\alpha$-abundance signature of its stars
(\citealt{robertson05,font06,johnston08}). Therefore, by studying
stars in the outer halo, we are studying stars from lower mass
satellites. An example of one of these possible abundance signatures
comes from comparing the abundances of nearby stars with stars in
current day dwarf spheroidals. For example, present day dSphs
sometimes show lower [$\alpha$/Fe] ratios relative to nearby disk and
halo stars (e.g, \citealt{shetrone01,venn04}).

To quantify changes in the stellar populations at large Galactocentric
radius, we have initiated a program to measure the abundances for a large sample
of metal-poor stars in the outer halo. Using the Sloan
Extension for Galactic Understanding and Exploration (SEGUE) survey to select
candidates, and the ESI instrument to obtain more detailed follow-up spectra, we
have discovered SDSS J234723.64+010833.4 to have anomalous [Mg/Fe] and
[Ca/Fe] abundance ratios, and to be unique (to our knowledge) in its
[Mg/Ca] ratio of $-1.21$.

\begin{figure}
\begin{center}
\scalebox{1}[1]{
\plotone{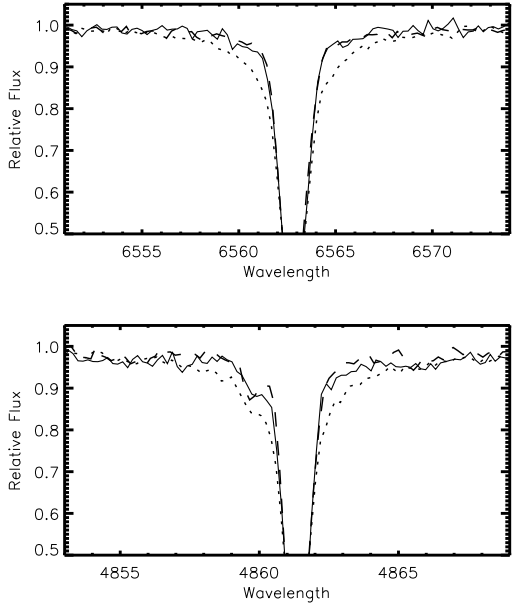}}
\end{center}
\figcaption{The H$\alpha$ and H$\beta$ lines for SDSS J234723.64+010833.4 (solid line), HD
  122563 (long dashed line), and CS 31078-018 (short dashed line). The
  Balmer lines from HD 122563 give a very good match to those measured
  in SDSS J234723.64+010833.4.
\label{balmer}}
\end{figure} 

\section{Observations and Reductions}

As part of the SEGUE survey, low-dispersion spectra have been obtained
for $\sim$240,000 stars, including stars specifically targeted for low
metallicity \citep{yanny09}.  Atmospheric parameters and metallicities
are derived from these spectra by the SEGUE Stellar Parameter Pipeline
(SSPP; \citealt{lee1,lee2,allende08}). From these stars, we have
selected a sample of stars with [Fe/H]$\leq -2.0$ and Galactocentric
distances $\geq 15$kpc (as determined from the surface gravity
estimate) for follow-up observations. For stars, this distant and faint
($V$ $\simeq 16.5$ and greater), a very efficient spectrograph is needed
to obtain a reasonable sample size. For this study, we use the ESI
spectrometer on Keck 2 \citep{sheinis02}. The 0''.75 slit was used,
which results in a resolving power of $R\sim6000$. The efficiency and
wavelength coverage of ESI (useable for detailed abundance analysis
between $\sim$4000 an 7000 \AA{}) makes this well suited to this
study, though the trade-off between efficiency and resolution limits
the number of elements that we can measure. It is in this sample that
we observed SDSS J234723.64+010833.4.

On 2008 August, 29 (UT) we obtained three 1500 s exposures of
SDSS J234723.64+010833.4 ($g=17.23$). We achieved a signal-to-noise
ratio (S/N) of 40 per pixel at 4170 \AA{}, increasing to 115 per pixel at
6200 \AA{}. The spectra were reduced and extracted
using the MAKEE
package.\footnote{http://spider.ipac.caltech.edu/staff/tab/makee/}
Equivalent widths were measured using the SPECTRE program \citep{spectre}.
The measured EWs along with line parameters are available upon
request (with atomic parameters taken from \citealt{lai08}). 

\section{Stellar Parameters and Analysis}

We follow the general treatment outlined in \citet{lai04,lai07} to
determine the atmospheric parameters and perform
the LTE spectral analysis for this star. We have
expanded the number of species measured to include \ion{Na}{1} and
\ion{Fe}{2}, and also use a \citet{castelli03} model atmosphere
with no convective overshooting.

We obtained initial atmospheric parameters of 5108 K for \teff{}, 2.15
for \logg{}, and [Fe/H] $=-2.2$ from the SSPP. In the course of the
analysis we found a large correlation between the excitation potential
(EP) and \ion{Fe}{1} abundances for individual \ion{Fe}{1} lines. We
need to adjust the SSPP temperature by approximately $-600$K to bring
this correlation below the 2-$\sigma$ level. However, this is a large
correction and is based on only 16 \ion{Fe}{1} lines. As an additional
check we compared the H$\alpha$ and H$\beta$ line profiles to stars
with better determined \teff{} and with ESI spectra from
\citet{lai04}. In Figure \ref{balmer} we show this comparison to HD
122563 and CS 31078-018, stars with similar \teff{} to the
EP-corrected \teff{} and SSPP \teff{} of SDSS J234723.64+010833.4,
respectively \citep{mashonkina, lai08}. The SDSS J234723.64+010833.4
Balmer line profiles very closely match those of HD 122563
(\teff{}$=4600$K), but are not a good fit to those of CS 31078-018
(\teff{}$=5257$K). At present, the origin of this rather large offset
in \teff{} remains unclear, but in examining our larger sample it
seems to only be present for some stars with \teff{}$<4900K$. The
metallicity offset is easier to understand. The high [Ca/Fe]
contributes to the offset in [Fe/H] with respect to the SSPP
value. Several of the metallicity indicators in the SSPP use the CaII K line
strength \citep{lee1}, so the unusually high [Ca/Fe] value would be
expected to bias the SSPP-adopted metallicity upward. Combined with
the \teff{} offset, this in large part explains the metallicity
difference.

This much lower \teff{} may also indicate the \logg{} from the SSPP
should be changed. Given the similarity to HD 122563 in \teff{}, one
reasonable choice is to adopt its \logg{} of 1.5 for SDSS
J234723.64+010833.4. As a check on the \logg{} of the star, we compare
the \ion{Fe}{1} to the \ion{Fe}{2} abundance. While we only measure
two \ion{Fe}{2} lines (at 4583.8 and 4923.93 \AA{}), we find that the
\ion{Fe}{1} and \ion{Fe}{2} abundances agree within reasonable errors
using the SSPP atmospheric parameters, but are completely inconsistent
when using the lower \teff{}. However, using the \logg{} of HD 122563
for SDSS J234723.64+010833.4 gives a much better agreement between the
two species.

We ultimately adopt the \teff{} of HD 122563, as given in
\citet{mashonkina} for SDSS J234723.64+010833.4, motivated by the
match in the H$\alpha$ and H$\beta$ lines of the two stars and
minimized EP trend of \ion{Fe}{1} lines. We also adopt the \logg{} of
HD 122563 predominately because of this \teff{} match, with the similar abundances
of \ion{Fe}{1} and \ion{Fe}{2} giving confidence to this choice. We
then use Equation (7) from \citet{kirby08a} to derive the microturbulent
velocity ($v_t$), which is based on \logg{}. Our final adopted
atmospheric parameters for [Fe/H], \teff{}, \logg{}, and $v_t$, are
$-3.2$, 4600 K, 1.50, and 1.94 km s$^{-1}$, respectively. Cross correlating
with a radial velocity standard, we measure a heliocentric radial
velocity of $-225$ km s$^{-1}$.

\subsection{Error estimates}

Although we adopt \teff{} of 4600 K, motivated both by reducing the
EP-abundance trend of \ion{Fe}{1} lines and the Balmer
line similarity to HD 122563, lowering the \teff{} by an additional 200 K would 
even better eliminate the EP-abundance trend in the \ion{Fe}{1}
lines. We therefore adopt 200 K as the error on \teff{}. We also adopt 0.5 dex as our error on \logg{}, as it is at this
level the condition of ionization balance between \ion{Fe}{1} and
\ion{Fe}{2} begins not to be well met (because of the paucity of
\ion{Fe}{2} lines, we define this to be a difference of 0.3 dex
between the two species). While the equation used for $v_t$ gives a
good approximation of typical values for a given \logg{}, we can
estimate its error based on the trend of abundances given by
individual \ion{Fe}{1} lines with EWs. The value given by the
\citet{kirby08a} equation yields no obvious trend for individual
\ion{Fe}{1} abundances with EW. Varying this value by 0.3 km s$^{-1}$ gives a
noticeable correlation, leading us to adopt this as the error on
$v_t$. We estimate the uncertainties from atomic parameters and EW
measurements with the standard error of the abundances measured from
multiple lines of the same species. When only one line of a certain
species is measured, we estimate a conservative error of 0.2 dex due
to these factors.

To estimate the final errors on our abundance ratios, we adopt the
technique described in \citet{johnson2002} (Equations (5) and (6)). This
includes cross-terms for the dependence of \teff{} and \logg{} (a
decrease of 200K would necessitate a decrease of $\sim$0.3 in \logg{}
to maintain reasonable ionization equilibrium) and one taking into
account the dependence of $v_t$ with \logg{}.

\begin{deluxetable}{lcrrr}
\tablecolumns{5}
\tabletypesize{\small}
\tablewidth{0pc}
\tablecaption{Abundance Summary \label{abundances}}
\tablehead{
       \colhead{Element} &  \colhead{[X/Fe]} &
       \colhead{$\sigma$} & \colhead{Number of} & \colhead{Total}\\ 
       \colhead{Name} & \colhead{} & \colhead{Lines} & \colhead{Lines}  & \colhead{Error}}
\startdata    
   Fe                 &  $-$3.17    &  0.18  & 16  & 0.21\\
   Fe{\scriptsize II} &  $-$3.18    &  0.11  & 2   & 0.20\\
   C                  &  $-$0.28    &  0.20  & \nodata & 0.26\\
   Na                 &  $-$0.23    &  \nodata & 1 & 0.22  \\
   Mg                 &  $-$0.10    &  0.24  & 4   & 0.15 \\
   Ca                 &     1.11    &  0.09  & 6   & 0.08 \\
   Ti{\scriptsize I} &     0.39     &   \nodata & 1 & 0.22 \\
   Ti{\scriptsize II} &    0.57\tablenotemark{*}   & 0.24 & 5 & 0.13 \\
   Cr                 &  $-$0.05  &  \nodata & 1 & 0.21 \\
   Sr{\scriptsize II} &  $<-0.50$\tablenotemark{*} & \nodata & \nodata & \nodata \\
   Ba{\scriptsize II} &  $<-1.50$\tablenotemark{*}  &  \nodata & \nodata & \nodata  \\
   Eu{\scriptsize II} &  $<-0.50$\tablenotemark{*} & \nodata & \nodata & \nodata
\enddata
\tablenotetext{*}{[X/Fe {\scriptsize II}].}
\end{deluxetable}

\section{Results}

The abundance measurements, number of lines measured for each element,
standard deviations of these lines, and final total error estimates are
summarized in Table \ref{abundances}. We note that if we used the
atmospheric parameters from the SSPP, the abundance ratios would
remain essentially the same, but the overall metallicity would be
shifted higher to [Fe/H]$\sim-2.6$. In particular, the [Mg/Fe] and
[Ca/Fe] ratios are relatively insensitive to changes in the
atmospheric parameters.

\begin{figure}
\begin{center}
\scalebox{1}[1]{
\plotone{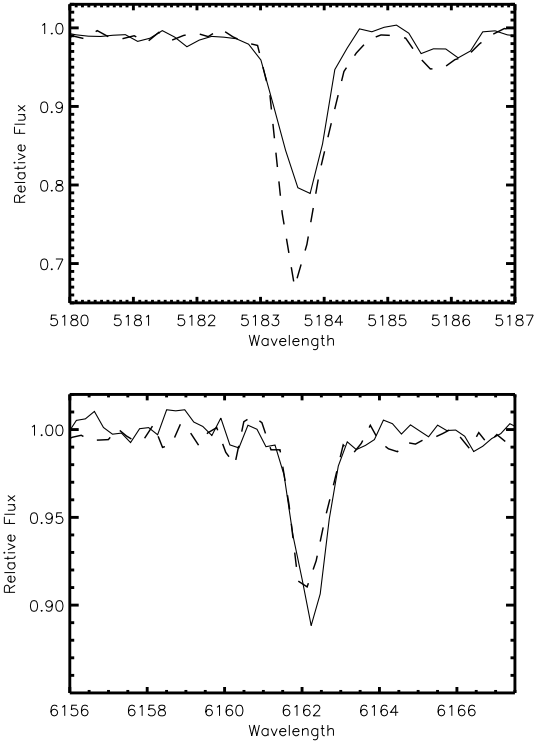}}
\end{center}
\figcaption{A Mg line at 5183 \AA{}(top plot) and a Ca line at 6162 \AA{}(bottom plot) from
  SDSS J234723.64+010833.4 (solid line), and HD
  122563 (long dashed line). HD 122563 has a [Fe/H]$=-2.6$ and
  relatively normal [$\alpha$/Fe]$\sim0.2$. SDSS J234723.64+010833.4,
  which has [Fe/H]$=-3.17$ should have weaker noticeably weaker lines
  if this were a typical [$\alpha$/Fe] enhanced MP star. However, while
  the Mg is clearly weaker, the Ca line is actually stronger relative
  to the same line in HD 122563.
\label{spectra}}
\end{figure} 

In Figure \ref{spectra}, we show a Mg and Ca line from our spectrum of
SDSS J234723.64+010833.4. We also plot the same lines from a spectrum
of HD 122563, which as discussed above should have very similar
atmospheric parameters. Even without additional analysis, we can see
that SDSS J234723.64+010833.4 is very enhanced in [Ca/Fe] and low in
[Mg/Ca] compared to HD 122563. The determination of the Mg abundance
comes from the 4703.0, 5172.7, 5183.6, and 5528.4 \AA{} lines, with
EWs ranging from 20.1 to 156.9 m\AA{}.The determination of the Ca
abundance comes from the 4226.7, 4425.4, 5588.8, 6122.2, 6162.2, and
6439.1 \AA{} lines, with EWs ranging from 76.9 to 348.5 m\AA{}. As can
be seen in Table \ref{abundances}, the individual lines of Mg and of Ca
are in good agreement with each other.

\begin{figure}
\begin{center}
\scalebox{1}[1]{
\plotone{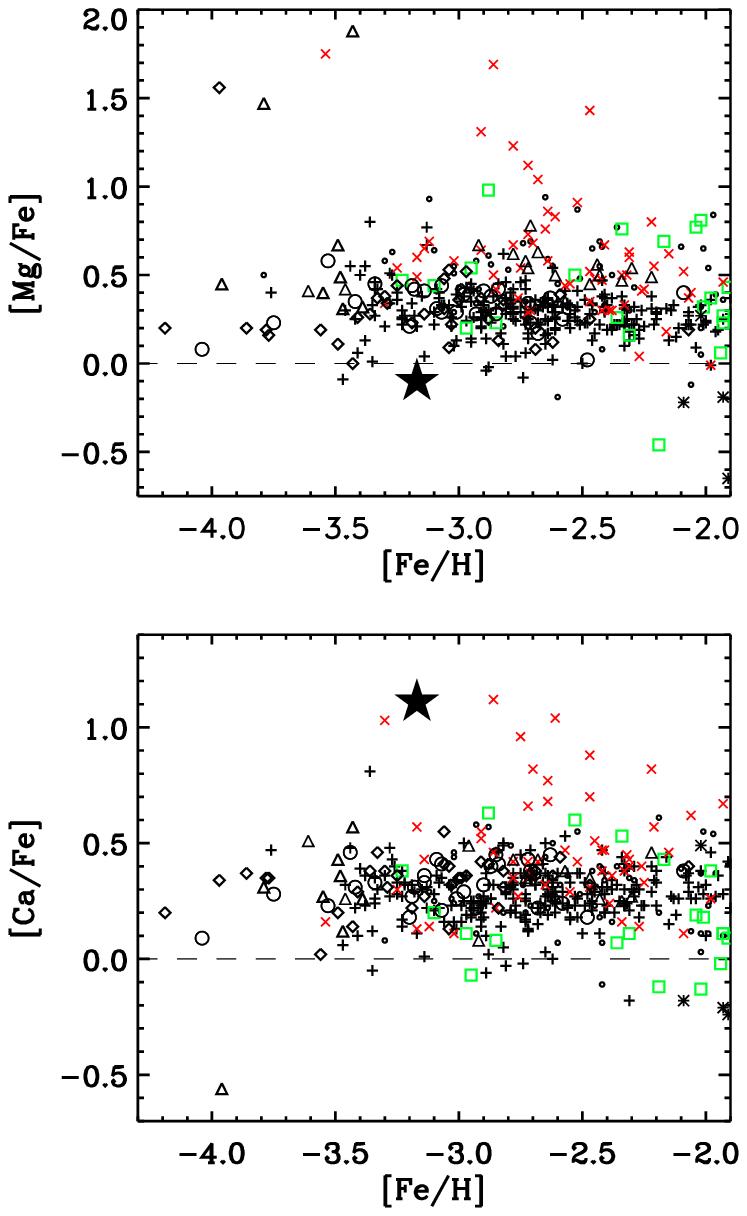}}
\end{center}
\figcaption{[Mg/Fe] and [Ca/Fe] for SDSS
  J234723.6+010833.4 (the filled star), along with data from multiple other
  recent studies. Most of the measurements are for nearby halo stars: the
  large circles are from \citet{lai08}, the x symbols are from
  \citet{aoki07,aoki07b} and references therein, the diamonds are from
  \citet{cayrel04}, the triangles are from \citet{cohen07-2,cohen04},
  the asterisks are from \citet{ivans03}, the plus
  signs are from \citet{heres2}, and the small circles are from
  \citet{lai04}. We have also included abundances measured in dSph
  galaxies. These are plotted as the squares and come from
  \citet{frebel09,koch08,fulbright04,shetrone03,shetrone01} (green in the
  electronic version). The only stars
  that seem to match the high [Ca/Fe] are the CEMP stars from the
  study of \citet{aoki07}.
\label{xfe}}
\end{figure}

\begin{figure}
\begin{center}
\scalebox{1}[1]{
\plotone{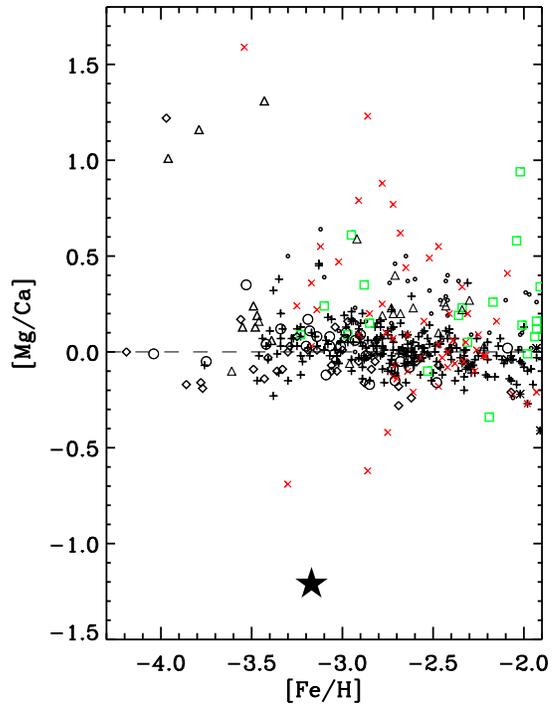}}
\end{center}
\figcaption{In terms of [Mg/Ca], SDSS J234723.64+010833.4 Is
  unique. The symbols are as in the previous figure. It is clear from
  this figure that most stars that exhibit anomalous
  $\alpha$-abundance ratios have either a normal or high [Mg/Ca], not
  the very low value exhibited by SDSS J234723.64+010833.4.
\label{mgca}}
\end{figure} 

In Figure \ref{xfe} we plot the [Mg/Fe] and [Ca/Fe] of SDSS
J234723.6+010833.4. The [Mg/Fe] is clearly lower than the vast
majority of stars; the only stars that show similarly high [Ca/Fe]
ratios come from the carbon-enhanced metal poor (CEMP) stars compiled
by \citet{aoki07}. However, \citet{aoki07} mentions that there are
numerous C$_2$ and CN features in the wavelength range of the Ca lines
measured in their highest [Ca/Fe] stars, and that this may be
artificially biasing the [Ca/Fe] values high in their CEMP
stars. Also, unlike the stars in \citet{aoki07}, SDSS
J234723.64+010833.4 has a very low carbon abundance, and, as shown in
Figure \ref{mgca}, its [Mg/Ca] rato of 1.21 $\pm 0.13$ is truly
different from other metal-poor stars. This value is 0.5 dex lower
then the next lowest star, and would be even more distinct if we
ignore the high [Ca/Fe] stars from \citet{aoki07}. As discussed
further below, most stars that exhibit anomalous $\alpha$-abundances
have similar [Mg/Fe] and [Ca/Fe], or high [Mg/Ca].

\section{Discussion}

The typical explanation for uniformly low or high $\alpha$-abundance
ratios in a star is that it came from a different stellar formation environment
compared to the nearby halo. Specifically, that the time-scale and
incidence of SN Ia versus SN II can account for the differences
(see, e.g., Figure 1 in \citealt{mcwilliam97}). This basic explanation can
also explain the $\alpha$ deficiencies found in current day dSphs
(e.g., \citealt{shetrone01,shetrone03,venn04}) and high space velocity
stars that may belong to the outer halo \citep{fulbright02,stephens02}. While the low
[Mg/Fe] of SDSS J234723.64+010833.4 may favor this scenario, the very
high [Ca/Fe] suggests that there is much more going on.

On the other end of the scale from SDSS J234723.64+010833.4 are the
high [Mg/Ca] stars. These seem to come in two types, stars with high
[Mg/Fe] ratios but normal [Ca/Fe] ratios \citep{aoki07b}, or low
[Ca/Fe] with normal [Mg/Fe] \citep{cohen07}. As discussed
in \citet{cohen07}, this behavior may arise because of the difference
in nucleosynthesis mechanisms between Mg and Ca, hydrostatic and
explosive $\alpha$-burning, respectively \citep{ww95}. This suggests
that in the progenitor(s) of SDSS J234723.6+010833.4, the products of
explosive nucleosynthesis dominated the ejecta as compared to the
products of hydrostatic burning. Measurements of Si (an explosive
$\alpha$-burning product) and O (a hydrostatic $\alpha$-burning
product) could greatly bolster this argument, however these two
elements have transitions that require bluer wavelength coverage and
higher-resolution spectra than ESI can provide.

The low [C/Fe] of this star is of particular note. Combined with the
low [Ba/Fe] upper limit, this indicates that the atmosphere has not
been polluted by a companion that has gone through its asymptotic
giant branch (AGB) phase. While it is possible that this star has processed some of its original
C to N, and therefore began with a higher original [C/Fe] abundance
ratio, it is unlikely that enough C was burned that this star was
ever a CEMP star ([C/Fe]$>$1). However, if somehow a significant
portion of its C has been converted to N internally, then some of its
Mg may also have been processed to Al through proton burning in layers
with T$\geq 70 \times 10^6$ K. If a mixing event that brought this C-depleted material to the surface was also deep enough to bring some of
this Mg-processed material to the surface, then the surface Mg
abundance would have been diluted and may partially explain the low
observed [Mg/Fe] value. While this is unlikely, in part because field stars do
not show a correlation of Mg with Al (e.g., \citealt{gratton04}) and
we measure a relatively low [Na/Fe] abundance ratio (which one would
expect to be enhanced in this scenario), further observations of this
star to determine its Al abundance can determine the validity of this
scenario.

If we assume that the low [Mg/Ca] was not the consequence of
processed material from a companion or internal to the star itself,
then this unique abundance ratio was present in the ISM that
originally formed the star. This indicates either a unique SN
progenitor or mix of progenitors giving rise to the [Mg/Ca] abundance
ratio. There do exist metal-free SN II models that can produce low [Mg/Ca]
ratios \citep{heger08}, and we find that these generally fall in between
the 10 and 18 $M_{\odot}$ range. However, because of the limited number
of abundances that can be measured from our spectrum, we are prevented
from performing these fits with confidence. Regardless, stars like
SDSS J234723.64+010833.4 must be exceedingly rare (at least in the
inner halo) given that this is the
first such star to display such a ratio. Clearly, this star is not a
product of the well-mixed ISM which seems to characterize the inner
halo. As such, it is an intriguing candidate for being the product of a
single unique progenitor and, combined with its metallicity, makes it
a possible true second-generation star \citep{tumlinson06-2}. A more
detailed high-resolution analysis of this star is needed before a more
definitive statement can be made.

An open question is whether this is a unique star formation
environment in the outer halo. Using its estimated surface gravity and
measured $g$ magnitude applied to the isochrones from
\citet{girardi04}, we estimate that SDSS J234723.64+010833.4 is located
approximately 40 kpc from the Galactic center (and $\sim32$ kpc below
the Galactic plane), well into the outer-halo region as defined from
the \citet{carollo07} measurements. In addition to the observational
evidence of a distinct outer-halo component, simulations have also
shown that an inner--outer stellar halo structure will arise assuming
the hierarchical merging picture (e.g, \citealt{delucia08,bullock05}),
with small satellites merging to form the outer halo. This
suggests that the outer stellar halo is composed of stars from many
different star-formation environments. This possibility is hinted at
by the study of \citet{roederer09}, which finds increasing chemical
abundance diversity for stars that have orbits that bring them to the
outer-halo region as compared to stars which solely exist in the inner
halo. 

While this star does not
resemble any abundances measured in dSphs to date, in the scenarios
described by the studies above it is possible that it once belonged to
a system that was accreted into the outer-halo region. The discovery
of more anomalous stars like SDSS J234723.64+010833.4 in the outer
halo could be a significant step in completing the picture for how the
stellar halo formed. It is noteworthy that this star was discovered in a sample of
only 27 in situ outer-halo stars. This suggests that, while obviously
rare, we may discover additional such stars as we increase our
sample of outer halo stars with available detailed abundance measurements.

\acknowledgements

Funding for the SDSS and SDSS-II has been provided by the Alfred
  P. Sloan Foundation, the Participating Institutions, the National
  Science Foundation, the U.S. Department of Energy, the National
  Aeronautics and Space Administration, the Japanese Monbukagakusho,
  the Max Planck Society, and the Higher Education Funding Council for
  England. The SDSS Web Site is http://www.sdss.org/.

  The authors wish to recognize and acknowledge the very significant
  cultural role and reverence that the summit of Mauna Kea has always
  had within the indigenous Hawaiian community.  We are most fortunate
  to have the opportunity to conduct observations from this
  mountain. We also acknowledge the anonymous referee for their
  careful reading and useful suggestions incorporated into the
  manuscript.

D.K.L. acknowledges the support of the National Science Foundation through the NSF Astronomy
and Astrophysics Postdoctoral Fellowship under award AST-0802292.

D.K.L., M.B., and J.A.J. acknowledge the support of the NSF under grants AST-0098617 and AST-0607770.

Y.S.L. and T.C.B. acknowledge partial support for this work from the NSF under
grants PHY 02-16783 and PHY 08-22648; Physics Frontier Center/Joint Institute
for Nuclear Astrophysics (JINA).

{\it Facilities:} \facility{Keck:II (ESI)}


\bibliography{all.bib}

\end{document}